\documentclass[a4paper,10pt]{article}

\usepackage[utf8x]{inputenc}

\usepackage{amsmath,amssymb,bm}
\usepackage{sidecap}
\usepackage{times}
\usepackage{braket}
\usepackage{ulem} 
\usepackage{graphicx,epsf}
\usepackage{float}
\usepackage[left=2.5cm,top=2.5cm,right=2.5cm,bottom=3cm,nohead,nofoot]{geometry}

\usepackage{bbold}
\usepackage{wrapfig}

\usepackage{natbib}

\usepackage{xcolor} 

\date{}              

\usepackage{fancyhdr}

\begin{document}

\begin{centering}
\LARGE\textbf{Spin-orbit interaction in square core-shell nanowires}
       
\vspace{12pt}      
\normalsize\textbf{Anna Sitek$^{\ast}$, Tudor Gabriel Dumitru$^{\dagger}$, Sigurdur I.\ Erlingsson$^{\dagger}$, and Andrei Manolescu$^{\dagger}$}

\vspace{0pt} 
\normalsize\textit{$^{\ast}$Institute of Theoretical Physics,  Wroclaw University of Science and Technology, 50-370 Wroclaw, Poland}\\ 
\normalsize\textit{$^{\dagger}$Department of Engineering, Reykjavik University, Menntavegur 1, IS-102 Reykjavik, Iceland}\\
\normalsize\textit{e-mail: anna.sitek@pwr.edu.pl}\\
\end{centering}
\vspace{6pt}

\noindent
\textbf{ABSTRACT}
\vspace{3pt}

We theoretically investigate the spin-orbit interaction of electrons confined in the outer regions of square core-shell nanowires. The polygonal cross section leads to the accumulation of low-energy electrons in the corners and the formation of a significant energy gap that separates these corner-localized states from higher-energy states localized along the sides.
We show that the low-energy states behave like the states of independent quantum wires, while the higher-energy states exhibit features characteristic of coupled wires.


\vspace{12pt}
\noindent
\textbf{1. INTRODUCTION}
\vspace{3pt}

Core-shell nanowires are radial heterostructures composed of a central quantum wire (core) surrounded by a layer of a different material deposited on the sides of the core. These structures exhibit a wide variety of features, making them promising building blocks for quantum nanodevices. One of their tunable properties is the electronic band structure. By selecting appropriate materials and designing the cross-sectional geometry, electrons can be confined within the core region \cite{Jadczak14}, or shifted to the outer shell region, effectively forming tubular conductors \cite{Blomers13, Sonner19}. The band alignment at the core-shell interface can be further engineered through controlled intermixing of the core and shell materials. This allows for a transition from a step-like potential profile in the case of clean shells to a tilted potential that penetrates deeply into the shell \cite{Fabian24}.

Due to the crystallographic properties of the constituent materials, core-shell nanowires typically exhibit polygonal cross sections, most commonly hexagonal \cite{Jadczak14, Blomers13,Sonner19}. However, triangular \cite{Qian04} and rectangular \cite{Guniat19, Fonseka19} core-shell geometries have also been realized. 
A characteristic feature of tubular structures with polygonal cross sections is the non-uniform, energy-dependent distribution of electrons around the circumference. In particular, low-energy electrons tend to localize in effective quantum wells formed at the corners, while higher-energy states shift toward the sides, forming an increasing number of maxima in the probability distributions with rising energy \cite{Sitek15, Ballester12}. These effects become more pronounced as the shell width and number of corners decrease \cite{Sitek15, Sitek16}. For sufficiently thin shells, low-energy electrons are depleted from the facets, resulting in the formation of well-separated one-dimensional channels along the sharp edges \cite{Ferrari09b}. Consequently, a single core-shell nanowire can behave as a system of multiple quantum wires and, for example, allow the formation of multiple Majorana states \cite{Manolescu17, Stanescu18}.

The formation of topological states \cite{Stanescu17} and the operation of many quantum devices are facilitated by spin-orbit interaction (SOI) \cite{Nadj_Perge10}, which couples the electron's spin to the magnetic field generated by the electron's own motion in an electrostatic field. In solid-state systems, SOI arises either from bulk inversion asymmetry (Dresselhaus) or from structural inversion asymmetry (Rashba). The latter is determined by the confining potential and can therefore be engineered during nanowire growth or externally tuned via electrostatic gating.

In this paper, we employ the $\bm{k}\!\cdot\!\bm{p}$ method to investigate the influence of Rashba-type spin-orbit interaction on the electronic spectra of electrons confined in the outer regions of square core-shell nanowires. In the following section, we introduce the model and methodology, Section 3 presents our results, final conclusions are summarized in Section 4.

\vspace{12pt}
\noindent
\textbf{2. METHODS}
\vspace{3pt}

We study the energies and spatial distributions of electrons confined within the shells of infinitely long nanowires. The systems under investigation feature an extended band-offset potential that linearly decreases from \(656.6\) meV to zero over a radial distance \(r\) from the core boundary. This potential arises due to intermixing between the core and shell materials. We employ the  \(\bm{k}\!\cdot\!\bm{p}\) method and apply the folding-down procedure \cite{Fabian07, Wojcik21} to derive the Hamiltonian describing conduction-band electrons under the influence of valence-band electrons,
\begin{equation}
\label{HSOI}
\mathcal{H}_{cc} =
\left\{\frac{\hbar^2}{2m^{*}}\left(  k_{z}^2  
-\frac{\partial^2}{\partial x^2}  
-\frac{\partial^2}{\partial y^2} \right)  
 + V_{\mathrm{BO}}(x,y) + V_{\mathrm{E}}(x,y)
\right\} \mathbb{1}_{2\times 2} + k_z \bigg(\alpha_x(x,y)\sigma_x + \alpha_y(x,y)\sigma_y\bigg)   \, ,
\end{equation}
where  \(m^*\) is the effective mass and \(k_z\) is the wave vector along  the nanowire growth direction. The potentials \(V_{\mathrm{BO}}(x,y)\) and \(V_{\mathrm{E}}(x,y)\) represent the extended band-offset potential and the potential due to the external electric field, respectively. 
The second term in Eq.~\ref{HSOI} accounts for the Rashba spin–orbit interaction (SOI), with Rashba coefficients given by:

\begin{subequations}
\label{alpha_xy}
\begin{minipage}{0.46\textwidth}
\begin{equation}	
   \alpha_x(x,y) = -\alpha_0 \frac{\partial}{\partial y}V(x,y)   
\end{equation}
\end{minipage}
\hspace{1.0 cm} and \hspace{0.0 cm}
\begin{minipage}{0.4\textwidth}
\begin{equation}
	 \alpha_y(x,y) =  \alpha_0 \frac{\partial}{\partial x}V(x,y) \, 
\end{equation}
\end{minipage}
\end{subequations} \\

\noindent
The amplitude \(\alpha_0\) depends on the shell material semiconductor band gap, spin-split gap and interband momentum matrix elements. Finally, \(\sigma_x\) and \(\sigma_y\) represent the Pauli matrices. 

We diagonalize the Hamiltonian using the finite-difference method on a polar grid. To model the square cross-section of the nanowire, we impose polygonal boundaries on the polar grid and retain only the points lying within these boundaries. The eigenstates of the Hamiltonian (Eq.~\ref{HSOI}) are then calculated under Dirichlet boundary conditions \cite{Sitek15}.

\vspace{12pt}
\noindent
\textbf{3. RESULTS} 
\vspace{3pt}

Below, we present results for an 8-nm-wide InAs shell deposited on a 57-nm-thick InP core. We investigate the spin–orbit interaction (SOI) arising from the extended band-offset potential, which preserves the nanowire symmetry, as well as the combined effect of this potential and symmetry breaking introduced by an external electric field.

\vspace{3pt}
\noindent
\textbf{3.1 Transverse states}
\vspace{3pt}

\begin{figure}[t]
	\centering
	\includegraphics[scale=0.67]{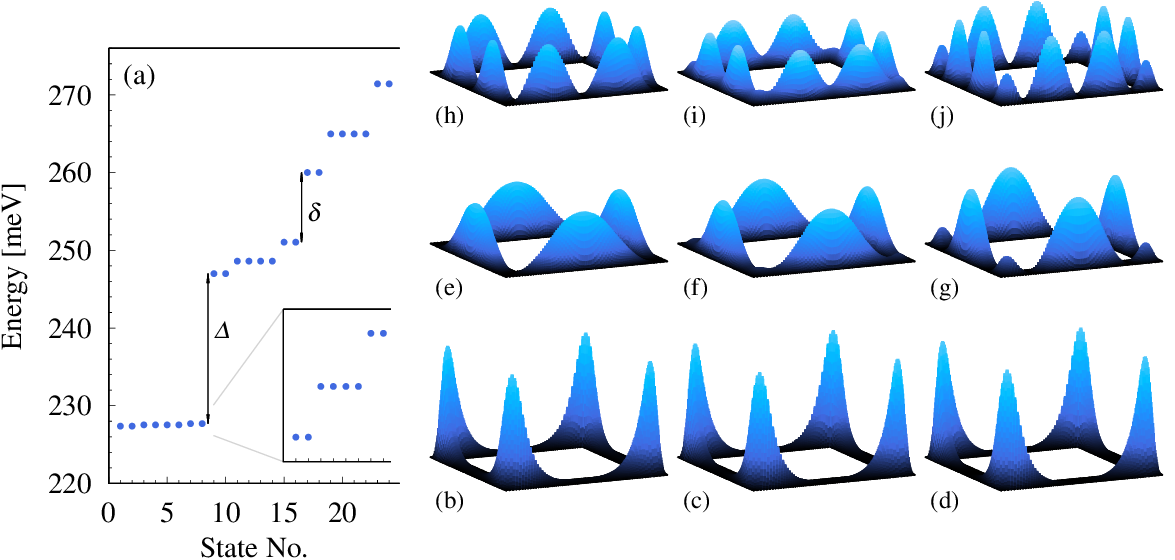}    
	\caption{(a) \textit{ Low-energy transverse states} ($\Delta = 19$ meV, $\delta = 9$ meV). (b)-(j) \textit{Electron probability distributions corresponding to the energy levels shown in panel} (a).}
	\label{fig:Ring}    
\end{figure}

The transverse energy states are the eigenstates of the cross sections of prismatic nanowires and effectively correspond to the energy levels of quantum rings. These states can be obtained by diagonalizing the Hamiltonian (Eq.~\ref{HSOI}) for \(k_z=0\). To focus on the properties of polygonal cross section, in this subsection we also set \(V_{\mathrm{BO}}(x,y)=0\) and \(V_{\mathrm{E}}(x,y)=0\).

The energy states of square quantum rings are grouped into sets of eight states—twice the number of corners. Within each group, the lowest and highest states are twofold degenerate due to spin only, while the middle states are fourfold degenerate due to a combination of spin and geometric symmetry [Fig.~\ref{fig:Ring}(a)]. The energy dispersion of each group increases with energy. Interestingly, the largest energy gap in the spectrum (\(\Delta\)) occurs between the two lowest groups of states. Although the separation between the second and third groups (\(\delta\)) is smaller, for the 8-nm ring it still exceeds the energy dispersion of the second group and remains significant.

Each group of eight states corresponds to a specific type of spatial localization. In particular, low-energy electrons are localized in the corner regions of the structure, Figs.\ \ref{fig:Ring}(b) - \ \ref{fig:Ring}(d). Electrons exited above the energy \(\Delta\) are shifted to the sides, the probability distributions associated with the second group of states form one maximum in every side,  Figs.\ \ref{fig:Ring}(e) - \ \ref{fig:Ring}(g), while those of the third group exhibit two maxima per side [Figs.~\ref{fig:Ring}(h)–\ref{fig:Ring}(j)]. These side-localized maxima may be accompanied by smaller peaks in the corners.

\vspace{3pt}
\noindent
\textbf{3.2 Spin-orbit interaction}
\vspace{3pt}

In the absence of SOI, the energy states of an infinite wire exhibit parabolic dispersions. For a square wire, each group of eight transverse states leads to the formation of three parabolic states. If the shell is clean, i.e., the contribution of the core material rapidly vanishes as the shell material begins to be deposited, then the electric field is non-zero only in a very narrow region around the core-shell interface. As a result, the impact of SOI on the energy bands is small. This effect can be significantly enhanced by increasing the intermixing length (\(r\)). The 
%
\begin{wrapfigure}{l}{0.38\textwidth}
\centering
\includegraphics[scale=0.67]{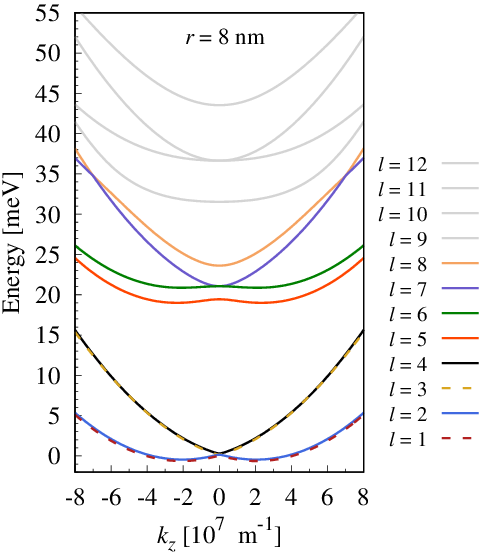}    
\caption{\textit{Energy dispersions in the presence of SOI induced by the extended band-offset potential.}}
\label{fig:SOI_r}    
\vspace{-0.3 cm}
\end{wrapfigure}
%
strongest SOI effect occurs when the contribution from core molecules gradually decreases throughout the entire shell, such that the extended band-offset potential reaches zero at the outer boundaries of the wire. In this case, the derivatives in Eqs.\ \ref{alpha_xy}, and thus the Rashba coefficients are non-zero across the entire cross-sectional area.
This potential does not break the square symmetry of the system, but effectively narrows the quantum well within the shell. The corresponding probability distributions differ from those shown in Fig.\ \ref{fig:Ring} only in minor details.

The SOI induced by the extended band-offset potential lifts only the fourfold degeneracies at finite wave vectors and shifts the corresponding states apart with increasing \(k_z\), Fig.\ \ref{fig:SOI_r}. As a result, all states become twofold degenerate for \(k\neq 0\). The wire under study is narrow enough to yield well-separated corner-localized maxima, resulting in a very small energy dispersion of the lowest energy states. In the presence of SOI, the states originating from the fourfold degenerate state at \(k_z=0\)  (\(l=2\) and \(l=3\)) nearly overlap with the upper and lower states, respectively. Consequently, the eight corner-localized energy states (from \(l=1\) to \(l=4\)) resemble the states of four nearly identical quantum wires.
In contrast, the side-localized states exhibit significantly larger energy dispersions. As a result, the four energy levels (from \(l=5\) to \(l=8\)) are grouped into two pairs of closely spaced but clearly separated states, which can interchange order at higher values of \(k_z\). The SOI has a qualitatively similar effect on the second group of corner-localized states (from \(l=9\) to \(l=12\)), though the impact is quantitatively smaller.

\begin{figure}[H]
\centering
\includegraphics[scale=0.67]{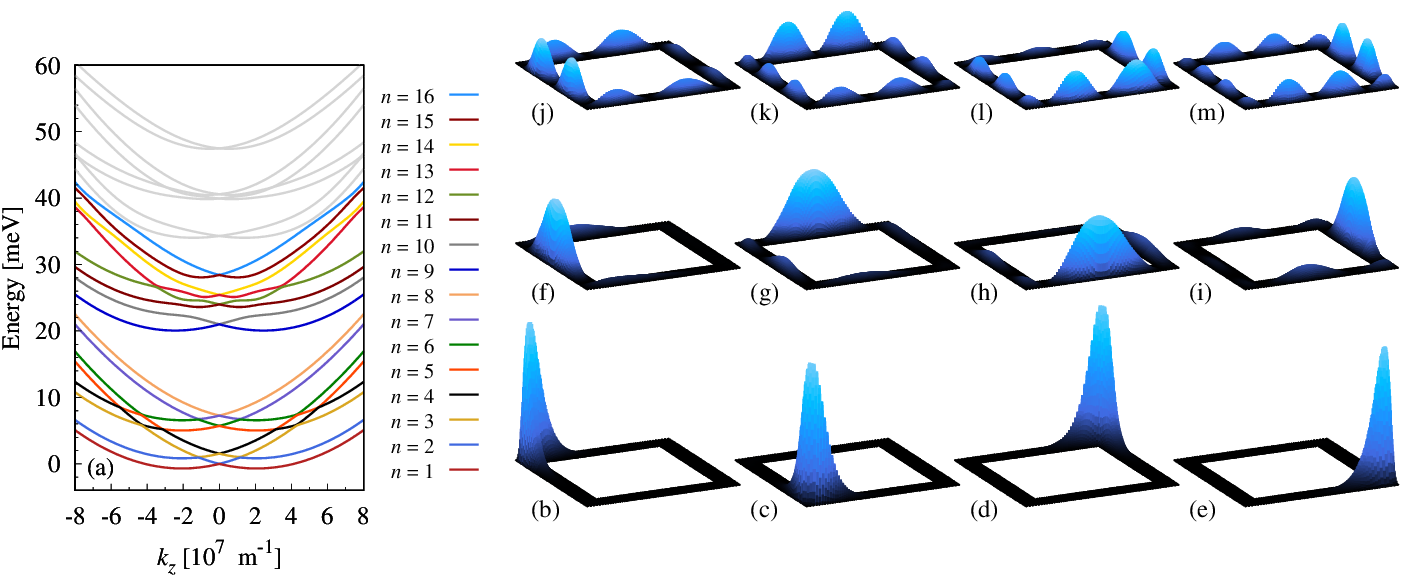}    
\caption{(a) \textit{ Energy dispersions in the presence of SOI due to the extended band-offset potential and the external electric field.} (b)-(m) \textit{ The corresponding probability distributions.} }
\label{fig:E_field_1}    
\end{figure}

The symmetry of the quantum wire can be broken by an external electric field. In Fig.\ \ref{fig:E_field_1}, the field is applied perpendicular to the wire and rotated by \(30^\circ\) with respect to the diagonal of the cross-section. This configuration breaks all possible symmetries of the square ring. As a result, the transverse energy levels become twofold degenerate due to spin only \cite{Sitek15}, meaning that each group of eight states is arranged into four distinct energy levels at \(k_z=0\). Each of these levels is associated with a different probability distribution.
In the presence of SOI, each twofold degenerate state splits into two branches, shifted along \(\pm k_z\) directions, leading to the formation of level crossings at \(k_z=0\).

The corner-localized states [from \(n=1\) to \(n=8\) in Fig.\ \ref{fig:E_field_1}(a)] resemble the energy spectrum of independent quantum wires with slightly different energies, i.e., they consist of shifted parabolic dispersions that either cross or exhibit minimal anti-crossings at finite \(k_z\). The corresponding probability distributions form a single sharp maximum in the corner regions, with the order of corner occupation determined by the orientation of the electric field, Figs.\ \ref{fig:E_field_1}(b)–\ref{fig:E_field_1}(e).
In contrast, the first group of side-localized states (from \(n=9\) to \(n=16\)) resembles the energy structure of four coupled quantum wires. These states form four pairs of parabolas that cross at \(k_z=0\) and 
%
%
\noindent
exhibit significant anti-crossings at finite  wave vectors. The associated probability distributions feature one dominant, elongated maximum, accompanied by smaller peaks, Figs.\ \ref{fig:E_field_1}(f)–\ref{fig:E_field_1}(i). These secondary maxima enable coupling between neighboring states.

The second group of corner-localized states [grey lines in Fig.\ \ref{fig:E_field_1}(a)] exhibits a much larger energy dispersion at \(k_z=0\) compared to the lower groups of states. However, the splitting between the two middle levels is minimal, consequently at finite \(k_z\) the middle states undergo a complex interaction. Their probability distributions, Figs.\ \ref{fig:E_field_1}(j)–\ref{fig:E_field_1}(m), are less sensitive to the electric field than those of the lower states and tend to form multiple maxima. Although the electric field reduces the energy gaps \(\Delta\) and \(\delta\), the splitting between corner- and side-localized states (between states \(n=8\) and \(n=9\)) remains sufficiently large to allow selective manipulation of the corner states.

\vspace{12pt}
\noindent
\textbf{4. CONCLUSIONS}
\vspace{3pt}

We investigated spin-orbit interaction in square core-shell nanowires, focusing on contributions from both the extended band-offset potential and an external electric field. In the presence of the extended band-offset potential alone, all energy states remain twofold degenerate. However, the application of an external electric field can lift all degeneracies at finite wave vectors. The lowest energy states in polygonal shells are localized along the sharp edges and are energetically separated from higher-energy states, which are localized on the facets. In the presence of SOI, the lowest states exhibit minimal anti-crossings or simple crossings at finite \(k_z\), indicating very weak or negligible interaction. In contrast, the first group of side-localized states displays significant anti-crossings, suggesting stronger coupling. As a result, a single core-shell nanowire can effectively behave as four non-interacting quantum wires at low energies and act as a system of coupled wires at higher energies.

\vspace{12pt}
\noindent
\textbf{ACKNOWLEDGEMENTS}
\vspace{3pt}

\noindent
This work was financed by the Icelandic Research Fund, Grant 195943, and by Reykjavik University Researcch Fund, Grant 223016.

\setlength{\bibsep}{0pt plus 0.3ex}
\bibliographystyle{plain}

\end{document}